# A Numerical Model for Flows in Porous and Open Domains Coupled at the Interface by Stress Jump

P. Yu[1,*], T.S. Lee[1], Y. Zeng[1], and H.T. Low[1,2]

[1]Department of Mechanical Engineering

[2]Division of Bioengineering,

National University of Singapore, Singapore 117576

**\* Email:**   g0202355@nus.edu.sg  or  xiaofishnus@hotmail.com

## Abstract:

A numerical model was developed for flows involving an interface between a homogenous fluid and a porous medium. The numerical model is based on the finite volume method with body-fitted and multi-block grids. The Darcy-Forchheimer extended model is used to govern the flow in the porous medium region. At its interface, a shear stress jump was imposed, together with a continuity of normal stress. Furthermore, the effect of the jump condition on the diffusive flux is considered, additional to that on the convective part which has been usually considered. Numerical results of three flow configurations are presented. The modeling is suitable for problems which have complex interface boundary conditions coupling between two flow domains.

## Key Words:

Interfacial condition, Stress jump, Porous medium, Block-structured grids

# INTRODUCTION

The study of flow systems which compose of a porous medium and a homogenous fluid has attracted much attention since they occur in a wide range of the industrial and environmental applications. Examples of practical applications are: flow past porous scaffolds in bioreactors, drying process, electronic cooling, ceramic processing, overland flow during rainfall, and ground-water pollution. Two different approaches, the single-domain approach [1, 2] and the two-domain approach [3, 4], are usually used to solve this type of problems.

In the single-domain approach, the composite region is considered as a continuum and one set of general governing equations is applied for the whole domain. The explicit formulation of boundary conditions is avoided at the interface and the transitions of the properties between the fluid and porous medium are achieved by certain artifacts [5]. Although this method is relatively easier to implement, the flow behavior at the interface may not be simulated properly, depending on how the code is structured [6].

In the two-domain approach, two sets of governing equations are applied to describe the flow in the two regions and additional boundary conditions are applied at the interface to close the two set of equations. This method is more reliable since it tries to simulate the flow behavior at the interface. Hence, in the present study, the two-domain approach, and the implementation of the interface boundary conditions, will be considered.

One of the several early studies on the interface boundary conditions is that by Beavers and Joseph [7]. In their approach, the fluids in a homogenous fluid and a porous medium are governed by the Navier-Stokes and Darcy equations respectively. A semi-empirical slip boundary-condition was proposed at the interface; because the flows in the different regions are governed by the corresponding partial differential equations of different orders. To make the governing equations of the same order, Neale and Nader [8] introduced a Brinkman term in the Darcy equation for the porous medium; and thus, proposed continuous boundary conditions in both stress and velocity. By matching both velocity and shear stress, Vafai and Kim [9] provided an exact solution for the fluid flow at the interface, which includes the inertia and boundary effects. However, a stress jump condition does occur at the interface as deduced by Ochoa-Tapia and Whitaker [10, 11] based on the non-local form of the volume averaged method.

Numerical solutions for the coupled viscous and porous flows have been attempted by many researchers [2, 3, 4, 12]. Different numerical methods such as finite volume method and finite element method have been used. Jue [2] simulated vortex shedding behind a porous square cylinder by finite element method. In his study, a general non-Darcy porous media model was applied to describe the flows both inside and outside the cylinder. A harmonic means was used to treat the sudden change between the fluid and porous medium. Costa *et al.* [4] proposed a control-volume finite element method to simulate the problems of coupled viscous and porous flows. A continuity of both velocity and stress at the interface was assumed and no special or additional procedure was needed to impose the interfacial boundary conditions.

However, different types of interfacial conditions between a porous medium and a homogenous fluid have been proposed; and found to have a pronounced effect on the velocity field as shown by Alazmi and Vafai [13]. Although the one-domain approach, or a continuity of both velocity and stress, is easier to implement, the more realistic stress jump condition has been adopted by many researchers.

The implementation of the numerical methodology on the stress jump condition can be found in the work of Silva and de Lemos [3]. Although they proposed that their treatment could be used in a complex geometry, their results were based on finite volume method in an orthogonal Cartesian coordinate system and for the case of fully developed flow. In their study, only the jump in shear stress was included and no special treatment on velocity derivatives was mentioned. However, for flow in general, it is needed to consider how to formulate the velocity derivatives at the interface. Also, for the two-dimensional problem, the normal stress condition is needed to close the sets of equations.

The objective of the present study was to develop a numerical model based on finite volume method to treat the stress jump condition given by Ochoa-Tapia and Whitaker [10, 11]. As the interface naturally divides the whole domain into different parts and its location is known *a priori*, the multi-block method is used. By combining body-fitted and multi-block grids, the method is effective for complex boundary conditions at the interface between different domains. The numerical model is more rigorous as it considers the effect of the stress jump condition on both convective and diffusive fluxes.

## MATHEMATICAL MODEL

Considering steady, laminar flow of an incompressible, viscous fluid, the governing equations for a homogenous fluid region, using vector form, can be written as:

$$\nabla \cdot \vec{u} = 0 \tag{1}$$

$$\nabla \cdot (\rho \vec{u}\vec{u}) = -\nabla p + \mu \nabla^2 \vec{u} \tag{2}$$

where $p$ is pressure, $\rho$ is mass density of the fluid, and $\mu$ is the fluid dynamic viscosity.

The porous medium is considered to be rigid, homogeneous and isotropic; and saturated with the same single-phase fluid as that in the homogenous fluid region. Considering viscous and inertia effects, the governing equations for porous region based on Darcy-Forchheimer extended model can be expressed as [14, 15]:

$$\nabla \cdot \langle \vec{u} \rangle = 0 \tag{3}$$

$$\nabla \cdot \left( \frac{\rho \langle \vec{u} \rangle \langle \vec{u} \rangle}{\varepsilon} \right) = -\nabla \left( \varepsilon \langle p \rangle^* \right) + \mu \nabla^2 \langle \vec{u} \rangle - \frac{\mu \varepsilon}{K} \langle \vec{u} \rangle - \frac{\rho \varepsilon C_F |\langle \vec{u} \rangle|}{\sqrt{K}} \langle \vec{u} \rangle \tag{4}$$

where the operators "$\langle \ \rangle$" and "$\langle \ \rangle^*$" identify the local average and the intrinsic average respectively; $\varepsilon$ is the porosity; $K$ is the permeability; and $C_F$ is Forchheimer coefficient. The Darcy velocity "$\langle \vec{u} \rangle$" and the intrinsic velocity "$\langle \vec{u} \rangle^*$" can be linked by the Dupuit-Forchheimer relationship, $\langle \vec{u} \rangle = \varepsilon \langle \vec{u} \rangle^*$.

At the interface between the homogeneous fluid and the porous medium, additional boundary conditions must be applied to couple the flows in the two regions. In the present study, the stress jump condition [10, 11] is applied:

$$\left. \frac{\mu}{\varepsilon} \frac{\partial \langle u \rangle_t}{\partial n} \right|_{porous\ medium} - \left. \mu \frac{\partial u_t}{\partial n} \right|_{homogeneous\ fluid} = \beta \frac{\mu}{\sqrt{K}} \langle u \rangle_t \Big|_{interface} \tag{5}$$

where $\langle u \rangle_t$ is the Darcy velocity component parallel to the interface aligned with the direction $t$ and normal to the direction $n$; $u_t$ is the fluid velocity component parallel to the interface; and $\beta$ is an adjustable parameter which accounts for the stress jump at the interface.

In addition to Equation (5), the continuity of velocity and normal stress prevailing at the interface are given by:

$$\vec{u}|_{\text{homogeneous fluid}} = \langle \vec{u} \rangle|_{\text{porous medium}} = \vec{v}_{\text{interface}} \tag{6}$$

$$\frac{\mu}{\varepsilon} \frac{\partial \langle u \rangle_n}{\partial n}\bigg|_{\text{porous medium}} - \mu \frac{\partial u_n}{\partial n}\bigg|_{\text{homogeneous fluid}} = 0 \tag{7}$$

where $\langle u \rangle_n$ is the Darcy velocity component normal to the interface; and $u_n$ is the fluid velocity component normal to the interface. By combining with the appropriate boundary conditions of the composite region, Equations (1) - (7) can be used to simulate the flow in system composed of a porous medium and a homogenous fluid.

## NUMERICAL MODEL

The finite volume method based on nonorthogonal grid is used to discretize the governing equations [16]. Cartesian velocity components are selected as dependent variables in the momentum equations; and the solution algorithm is based on SIMPLEC method. The non-staggered grid arrangement is applied and the Rhie and Chow interpolation [17] is used to obtain a suitable coupling between pressure and velocity.

In some cases, the location of the interface between a porous medium and a homogenous fluid is known *a priori*. This interface naturally divides the composite region into different parts. Also, in some cases, structured grids are difficult, even impossible, to construct for complex geometries. Therefore, in present study, block-structured grids method is applied.

There are in general three types of interfaces when the block-structured grids method is employed to calculate the flow in the composite region: fluid - fluid interface, porous medium - porous medium interface, and homogenous fluid - porous medium interface.

A typical control volume is shown in Figure 1. For a general dependent variable $\varphi$, a final discrete form over the control volume can be written as:

$$F_e + F_w + F_n + F_s = S \tag{8}$$

where $F_e$, $F_w$, $F_n$ and $F_s$ are the overall fluxes (including both convection and diffusion) of $\varphi$ at faces *e, w, n, s*, which denote *east, west, north,* and *south* of the control volume; and *S* the source term. The detailed numerical methodology for obtaining the convective flux ($F_e^c$, $F_w^c$, $F_n^c$, and $F_s^c$) and diffusive flux ($F_e^d$, $F_w^d$, $F_n^d$, and $F_s^d$) can be found elsewhere [16] and will only be outlined here.

With the midpoint rule approximation, the convective flux at face *east* can be calculated as:

$$F_e^c = \int_{S_e} \rho \varphi \vec{u} \cdot \vec{n} dS \approx m_e \varphi_e \tag{9}$$

where $m_e = \rho_e \left( S^x u + S^y v \right)_e$ and $\varphi_e$ is the value of $\varphi$ at the center of the cell face. To avoid the non-orthogonal effect, the midpoint rule with the deferred correction term [18] applied to the integrated diffusive flux gives:

$$F_e^d = \mu_e \left( \frac{\partial \varphi}{\partial \mathrm{n}} \right)_e S_e = \mu_e S_e \left( \frac{\partial \varphi}{\partial \xi} \right)_e + \mu_e S_e \left[ \overline{\left( \frac{\partial \varphi}{\partial \mathrm{n}} \right)_e} - \overline{\left( \frac{\partial \varphi}{\partial \xi} \right)_e} \right]^{old} \tag{10}$$

An implicit flux approximation of the term $\left(\dfrac{\partial \varphi}{\partial \xi}\right)_e$ is applied:

$$\left(\dfrac{\partial \varphi}{\partial \xi}\right)_e = \dfrac{\varphi_E - \varphi_P}{L_{PE}} \tag{11}$$

where $L_{PE}$ stands for the distance between nodes $P$ and $E$. The deferred correction terms can be obtained as:

$$\overline{\left(\dfrac{\partial \varphi}{\partial n}\right)_e} = \overline{(grad\varphi)_e} \bullet \vec{n}; \quad \overline{\left(\dfrac{\partial \varphi}{\partial \xi}\right)_e} = \overline{(grad\varphi)_e} \bullet \vec{i}_\xi \tag{12}$$

where $\vec{i}_\xi$ is the unit vector in the $\xi$-direction. The final expression of Equation 10 then becomes:

$$F_e^d = \mu_e S_e \dfrac{\varphi_E - \varphi_P}{L_{PE}} + \mu_e S_e \overline{(grad\varphi)_e}^{old} \bullet (\vec{n} - \vec{i}_\xi) \tag{13}$$

To obtain the deferred derivatives at the cell face, the derivatives are calculated first at the control volume centers and then interpolated to the cell faces. By using the Gauss' theorem, the derivative at the CV centers can be approximated by the average value over the cell:

$$\left(\dfrac{\partial \varphi}{\partial x_i}\right)_P \approx \dfrac{\int_\Omega \dfrac{\partial \varphi}{\partial x_i} d\Omega}{\Delta \Omega} = \int_S \varphi \vec{i}_i \bullet \vec{n} dS \approx \sum_c \varphi_c S_c^i, \quad c = e, n, w, s \tag{14}$$

The different methods to approximate the value of $\varphi$ and its derivative at the cell face result in different interpolation schemes. In present study, a second order scheme, the central difference scheme (CDS) is used.

For the interfaces treatment, the fluid – fluid interface and the porous medium – porous medium interface are easier to be implemented because there are no sudden changes of properties near the interface region. In this case, the two types of block interfaces are treated as the interior cell faces rather than boundaries, which was proposed by Lilek *et al.* [19]. The grids in two neighboring blocks match at the interface in present study.

The present model also considers the interface between a porous medium and a homogeneous fluid as the interior cell face. However, the calculation of convective and diffusive fluxes at the interface has to be reformulated because the stress jump condition is used and the linear interpolation (CDS) fails to approximate the fluxes there. The formulation of the convective and diffusive fluxes at the interface is developed from Equations (5) to (7).

Figure 2 shows details of the interface between a porous medium and a homogeneous fluid. Two neighbor control volumes, lying in the homogenous fluid and the porous region respectively, share the interface. The velocity vector at the interface is given by $\vec{v}_{interface}$. It can be written in either the *x-y* or *n-t* coordinate systems as

$$\vec{v}_{interface} = \langle u \rangle_x \vec{e}_x + \langle u \rangle_y \vec{e}_y = \langle u \rangle_n \vec{n} + \langle u \rangle_t \vec{t} \qquad (15)$$

where $\langle u \rangle_x$ and $\langle u \rangle_y$ are the components of $\vec{v}_{interface}$ in the *x* and *y* directions while $\langle u \rangle_n$ and $\langle u \rangle_t$ are the $\vec{v}_{interface}$ components along *n* and *t* directions respectively. And the component $\langle u \rangle_t$ then can be written as:

$$\langle u \rangle_t = \langle u \rangle_x \vec{e}_x \cdot \vec{t} + \langle u \rangle_y \vec{e}_y \cdot \vec{t} \qquad (16)$$

By combining Equations (5), (7) and (15):

$$\frac{\mu}{\varepsilon} \frac{\partial \vec{v}_{interface}}{\partial n} \bigg|_{porous\ medium} - \mu \frac{\partial \vec{v}_{interface}}{\partial n} \bigg|_{homogeneous\ fluid} = \beta \frac{\mu}{\sqrt{K}} \langle u \rangle_t \vec{t} \qquad (17)$$

The unit vector ($\vec{t}$) parallel to the interface (Figure 2) is calculated from:

$$\vec{t} = \frac{(x_{ne} - x_{se})\vec{e}_x + (y_{ne} - y_{se})\vec{e}_y}{\sqrt{(x_{ne} - x_{se})^2 + (y_{ne} - y_{se})^2}} = \frac{\Delta x_e \vec{e}_x + \Delta y_e \vec{e}_y}{l_e} \qquad (18)$$

By substituting the components of $\vec{v}_{\text{interface}}$ in the *x* and *y* directions, the Equation (16) becomes:

$$\frac{\mu}{\varepsilon}\frac{\partial \langle u \rangle_x}{\partial n}\bigg|_{\text{porous medium}} - \mu \frac{\partial u_x}{\partial n}\bigg|_{\text{homogeneous fluid}} = \beta \frac{\mu}{\sqrt{K}} \frac{\langle u \rangle_x \Delta x_e \Delta x_e + \langle u \rangle_y \Delta y_e \Delta x_e}{l_e^2} \qquad (19)$$

$$\frac{\mu}{\varepsilon}\frac{\partial \langle u \rangle_y}{\partial n}\bigg|_{\text{porous medium}} - \mu \frac{\partial u_y}{\partial n}\bigg|_{\text{homogeneous fluid}} = \beta \frac{\mu}{\sqrt{K}} \frac{\langle u \rangle_x \Delta x_e \Delta y_e + \langle u \rangle_y \Delta y_e \Delta y_e}{l_e^2} \qquad (20)$$

The derivatives at the interface are calculated from the values at auxiliary nodes $P'$ and $E'$; these nodes lie at the intersection of the cell face normal n and straight lines connecting nodes *P* and *N* or *E* and *NE*, respectively, as shown in Figure 2. The normal gradients at the interface can be calculated by using the first order difference approximation:

$$\frac{\partial \langle u \rangle_x}{\partial n}\bigg|_{\text{porous medium}} = \frac{\langle u \rangle_x|_{E'} - \langle u \rangle_x|_e}{L_{eE'}}, \quad \frac{\partial \langle u \rangle_y}{\partial n}\bigg|_{\text{porous medium}} = \frac{\langle u \rangle_y|_{E'} - \langle u \rangle_y|_e}{L_{eE'}} \qquad (21)$$

$$\frac{\partial u_x}{\partial n}\bigg|_{\text{homogeneous fluid}} = \frac{u_x|_e - u_x|_{P'}}{L_{P'e}}, \quad \frac{\partial u_y}{\partial n}\bigg|_{\text{homogeneous fluid}} = \frac{u_y|_e - u_y|_{P'}}{L_{P'e}} \qquad (22)$$

The Cartesian velocity components at $P'$ and $E'$ can be calculated by using bilinear interpolation or by using the gradient at the control volume center:

$$u_x|_{P'} = u_x|_P + (grad u_x)_P \cdot \overrightarrow{P'P} \qquad (23)$$

To obtain higher order approximation of the derivatives, the velocity components at more auxiliary nodes may be needed. Alternatively, the shape functions may be used, which produces a kind of combined Finite Element/Finite Volume method for calculating the higher order approximations.

By making use of Equations (19) to (23), the Cartesian velocity components $\langle u \rangle_x$ and $\langle u \rangle_y$ at the interface are obtained. Then the convective fluxes at the interface can be calculated. The diffusive fluxes are calculated from Equations (21) - (23). By substituting the fluxes in Equation (8), and solving the resultant algebraic equations, the flow field can be obtained in all domains.

## RESULTS AND DISCUSSION

The numerical results of three flow situations will be presented: flow in a channel partially filled with a layer of a porous medium, flow through a channel with a porous plug, and flow around a square porous cylinder. All the results presented are grid-independent.

### 1. Flow in a channel partially filled with a layer of a porous medium

The physical domain is shown schematically in Figure 3. It consists of a planar channel which is horizontally divided into a homogenous fluid region with height $H_1$ above and a fluid-saturated porous region with height $H_2$ below. The case of height ratio $H_2/H_1 = 1$ is considered.

The flow is assumed laminar and fully developed. The governing equations are simplified as follows:

$$\frac{d^2u}{dy^2} = \frac{1}{\mu}\frac{dp}{dx} \qquad \text{for homogenous fluid} \qquad (24)$$

$$\frac{d}{dy}\left(\frac{\mu}{\varepsilon}\frac{d\langle u \rangle}{dy}\right) = \frac{1}{\varepsilon}\frac{d(\varepsilon p_f)}{dx} + \frac{\mu}{K}\langle u \rangle + \frac{\rho C_F}{\sqrt{K}}\langle u \rangle^2 \qquad \text{for porous medium} \qquad (25)$$

Introducing the dimensionless variables

$$U = \frac{\mu u}{GH_1^2} \text{ and } Y = \frac{y}{H_1}, \text{ where } G = -\frac{dp_f}{dx}$$

Equations (24) and (25) can be rewritten as:

$$\frac{d^2U}{dY^2} = -1 \qquad \text{for the homogenous fluid} \qquad (26)$$

$$\frac{1}{\varepsilon}\frac{d^2\langle U \rangle}{dY^2} = -1 + \frac{1}{Da}\langle U \rangle + F\langle U \rangle^2 \qquad \text{for porous medium region} \qquad (27)$$

where Darcy number $Da = \frac{K}{H_1^2}$ and Forchheimer number $F = \frac{C_F \rho G H_1^4}{K^{1/2}\mu^2}$. The boundary conditions are:

$$U = 0 \text{ at } Y = 1 \text{ and } \langle U \rangle = 0 \ Y = -H_2/H_1 \qquad (28)$$

$$\frac{1}{\varepsilon}\frac{d\langle U \rangle}{dY} - \frac{dU}{dY} = \frac{\beta_1}{\sqrt{Da}}U_{\text{interface}} \text{ at } Y = 0 \qquad (29)$$

Following the proposal of Nield *et al.* [20], Equations (26) to (29) can be solved analytically as shown in the Appendix. Both numerical and analytical solutions are presented for validation of the present numerical implementation.

Figure 4 shows the *u* velocity profile under different flow conditions. It is seen that the numerical and analytical results are in good agreement. The effect of the Darcy number (*Da*) on the *u* velocity profile is presented in Figure 4a in which the *Da* varies from $10^{-3}$ to $10^{-1}$ while the other parameters are kept constant. Although in practical applications, *Da* may not go up to $10^{-2}$ [21], nevertheless this range is also presented to show *Da* effect more clearly. From Figure 4a, it is seen that the *u* velocity decreases with the increase of *Da*. When *Da* is less than $10^{-3}$, the *u* velocity in the porous medium is almost zero except the region near the interface. The effect of the porosity ($\varepsilon$) on the u profile is shown in Figure 4b. The *u* velocity should decrease as the porosity decreases and the numerical and analytical results seem to show the trend. However, in the porous medium region around -0.33 < *Y* < -0.62, the *u* velocity is slightly larger when the porosity is smaller. This may be because the $\beta$ chosen here is kept constant even though it should vary with the variation of $\varepsilon$. The Forchheimer number (*F*) does not have much effect on the velocity distribution. The velocity decreases slightly when *F* increases from 1 to 100 as shown in Figure 4c. The effect of the jump parameter ($\beta$) on the flow is shown in Figure 4d, which indicates that the u velocity increases noticeably as $\beta$ increases.

## 2. Flow through a channel with a porous plug

The physical domain of the second problem is shown schematically in Figure 5, which is the same as that by Gartling *et al.* [12] and Costa *et al.* [4]. In this problem the flow passes through a planar channel with a porous plug under an imposed overall pressure gradient. Different from the first problem, the governing dimensionless parameters are: Reynolds number based on the mean velocity, $Re = \rho UH / \mu$, Darcy number $Da = \dfrac{K}{H^2}$, the porosity $\varepsilon$, Forchheimer coefficient $C_F$ and jump parameter $\beta$. In present study, these parameters are chosen as $Re = 1$, $Da = 10^{-2}$, $\varepsilon = 0.7$, $C_F = 1$ and $\beta = 0.7$.

The numerical results are shown in Figure 6, where the centerline u velocity and pressure along *x* direction are presented. It is seen that the velocity drops rapidly in the porous plug, across which there is a large pressure drop. The flow field is predominantly axial over most of the homogenous fluid and porous medium regions, but it is two-dimensional in the region near the interface between the homogenous fluid and the porous medium. The trends of the present results are in general agreement with those of Gartling *et al.* [12] and Costa *et al.* [4]. However, there are slight differences in the velocity magnitude in the porous medium are because the stress jump conditions are applied in the present study whereas the continuous stress conditions are applied previously.

## 3. Flow around a square porous cylinder

The above two problems concern internal flow problems with regions of a homogenous fluid and a porous medium. To illustrate an external flow problem with complex geometry, the flow around a square porous cylinder is considered. The computational domain and mesh are shown schematically in Figure 7 and Figure 8 respectively. The governing dimensionless parameters are the same as those of the porous plug problem above: Reynolds number based on the mean velocity and the height of the cylinder, $Re = \rho U H / \mu$, Darcy number $Da = \dfrac{K}{H^2}$, the porosity $\varepsilon$, Forchheimer coefficient $C_F$ and jump parameter $\beta$. In this study, $Re$ is chosen as $Re = 20$ to ensure the steady and laminar flow.

Three different $Da = 10^{-2}$, $10^{-3}$ and $10^{-4}$ are chosen and other parameters are kept constant namely, $\varepsilon = 0.4$, $C_F = 1$ and $\beta = 0.7$.

The flow streamline for the flow around a square porous cylinder at different $Da$ are presented in Figure 9. At smaller $Da$ (=$10^{-4}$ in Figure 9a), that is the cylinder permeability is small, very little fluid flows through the cylinder. Hence the flow field resembles that around a solid cylinder. When $Da$ increases to $10^{-3}$ (Figure 9b), the vortex in the wake is reduced as there is more bleed fluid. At higher $Da$ (=$10^{-2}$ in Figure 9c), the large bleed flow has prevented vortex formation.

## CONCLUDING REMARKS

A general numerical model was developed for problems with interface between a homogenous fluid and a porous medium. The numerical model is based on finite volume method with body-fitted and multi-block grids, which is effective for dealing with complex boundary conditions at the interface between different domains. The shear stress jump condition is applied at the interface, and affects both the convective and diffusive fluxes. The normal stress condition, assumed continuous at the interface, is also needed in order to close the two sets of equations. The general model was applied to solve three flow configurations.

The numerical model of flow over a porous layer extends the work of Silva and Lemos [3] by using multi-block grid and including the diffusive flux term. The results compares well with the analytical solution. The numerical results for flow through a porous plug exhibit slight difference in velocity in the porous medium, as compared with the studies of Costa et al. [4] and Gartling et al. [12] which assume continuity of shear and normal stresses at the interface. The numerical model of flow past a porous cylinder includes coupling of two domains (fluid and porous medium); it continues on the study of Jue [2] which, however, was based on a one-domain approach.

# APPENDIX

Integrating Equation (27) yields:

$$\frac{1}{\varepsilon}\left(\frac{d\langle U\rangle}{dY}\right)^2 = \frac{2}{3}F\langle U\rangle^3 + \frac{\langle U\rangle^2}{Da} - 2\langle U\rangle + C \tag{A1}$$

Making use of Equations (28) and (29), the constant C in Equation (A1) can be yielded:

$$C = \varepsilon\left[\frac{\beta_1}{\sqrt{Da}}U_i + \left(\frac{1}{2} - U_i\right)\right]^2 - \left(\frac{2}{3}FU_i^3 + \frac{U_i^2}{Da} - 2U_i\right) \tag{A2}$$

Since $\dfrac{d\langle U\rangle}{dY}$ should be real and negative at this region, we can obtain:

$$\frac{d\langle U\rangle}{dY} = -\sqrt{\varepsilon\left(\frac{2}{3}F\langle U\rangle^3 + \frac{\langle U\rangle^2}{Da} - 2\langle U\rangle + C\right)} \tag{A3}$$

Then we can integrate Equation A3 to obtain:

$$\int_0^{U_{\text{interface}}} Q(\langle U\rangle)d\langle U\rangle = -\frac{H_2}{H_1} \tag{A4}$$

where $Q(\langle U\rangle) = -\left[\varepsilon\left(\dfrac{2}{3}F\langle U\rangle^3 + \dfrac{\langle U\rangle^2}{Da} - 2\langle U\rangle + C\right)\right]^{-1/2}$

In a similar fashion, we can obtain:

$$\int_{\langle U\rangle}^{U_{\text{interface}}} Q(\langle U\rangle)d\langle U\rangle = Y \tag{A5}$$

Given the values of *Da*, *F* and *ε*, the value of $U_{\text{interface}}$ is given in an inverse fashion by Equation (A4). Pairs of values (*Y*, $U_i$) determining the velocity profile can then be obtained from Equation (A5). Then the integrals in Equation (A4) can be solved using Romberg's numerical integration method.

# REFERENCES

1. Mercier J, Weisman C, Firdaouss M and Quéré PL, Heat Transfer Associated to Natural Convection Flow in a Partly Porous Cavity, *ASME J. Heat Transfer*, vol. 124, pp. 130-143, 2002.

2. Jue TC, Numerical Analysis of Vortex Shedding Behind a Porous Cylinder, *Int. J. Numer. Methods Heat Fluid Flow*, vol. 14, pp. 649-663, 2004.

3. Silva RA and de Lemos MJS, Numerical Analysis of the Stress Jump Interface Condition for Laminar Flow Over a Porous Layer, *Numer. Heat Transfer A*, vol. 43, pp. 603-617, 2003.

4. Costa VAF, Oliveira LA, Baliga BR and Sousa ACM, Simulation of Coupled Flows in Adjacent Porous and Open Domains Using a Control-Volume Finite-Element Method, *Numer. Heat Transfer A*, vol. 45, pp. 675-697, 2004.

5. Goyeau B, Lhuillier D, Gobin D and Velarde MG, Momentum Transport at a Fluid-Porous Interface, *Int. J. Heat Mass Transfer*, vol. 46, pp. 4071-4081, 2003.

6. Nield DA, Discussion, *ASME J. Heat Transfer*, vol. 119, pp. 193-194, 1997.

7. Beavers GS and Joseph DD, Boundary Conditions at a Natural Permeable Wall, *J. Fluid Mech.*, vol. 30, pp. 197-207, 1967.

8. Neale G and Nader W, Practical Significance of Brinkman's Extension of Darcy's Law: Coupled Parallel Flows within a Channel and a Bounding Porous Medium, *Can. J. Chem. Engrg.*, vol. 52, pp. 475-478, 1974.

9. Vafai K and Kim SJ, Fluid Mechanics of the Interface Region between a Porous Medium and a Fluid Layer – an exact solution, *Int. J. Heat Fluid Flow*, vol. 11, pp. 254-256, 1990.

10. Ochoa-Tapia JA and Whitaker S, Momentum Transfer at the Boundary between a Porous Medium and a Homogeneous Fluid I: Theoretical Development, *Int. J. Heat Mass Transfer*, vol. 38, pp. 2635-2646, 1995.

11. Ochoa-Tapia JA and Whitaker S, Momentum Transfer at the Boundary between a Porous Medium and a Homogeneous Fluid II: Comparison with Experiment, *Int. J. Heat Mass Transfer*, vol. 38, pp. 2647-2655, 1995.

12. Gartling DK, Hickox CE and Givler RC, Simulation of Coupled Viscous and Porous Flow Problems, *Comp. Fluid Dyn.*, vol. 7, pp. 23-48, 1996.

13. Alazmi B and Vafai K, Analysis of Fluid Flow and Heat Transfer Interfacial Conditions between a Porous Medium and a Fluid Layer, *Int. J. Heat Mass Transfer*, vol. 44, pp. 1735-1749, 2001.

14. Hsu CT and Cheng P, Thermal Dispersion in a Porous Medium, *Int. J. Heat Mass Transfer*, vol. 33, pp. 1587-1597, 1990.

15. Nithiarasu P, Seetharamu KN and Sundararajan T, Finite Element Modelling of Flow, Heat and Mass Transfer in Fluid Saturated Porous Media, *Arch. Comput. Meth. Engng.*, vol. 9, pp. 3-42, 2002.

16. Ferziger JH and Perić M, Computational Methods for Fluid Dynamics, 2$^{nd}$ ed., pp. 222-233, Springer, Berlin, 1999.

17. Rhie CM and Chow WL, Numerical Study of the Turbulent Flow Past an Airfoil with Trailing Edge Separation, *AAIA J.*, vol. 21, pp. 1525-1532, 1983.

18. Muzaferija S, Adapative Finite Volume Method for Flow Predictions Using Unstructured Meshes and Multigrid Approach. PhD Thesis, University of London, 1994.

19. Lilek Ž, Muzaferija S, Perić M and Seidl V, An Implicit Finite-Volume Method Using Nonmatching Blocks of Structured Grid, *Numer. Heat Transfer B*, vol. 32, pp. 385-401, 1997.

20. Nield DA, Junqueira SLM. and Lage JL, Forced Convection in a Fluid-Saturated Porous-Medium Channel with Isothermal or Isoflux Boundaries, *J. Fluid Mech.* Vol. 322, pp. 201-214, 1996.

21. Large JL, Effect of the Convective Inertia Term on Bénard Convection in a Porous Medium, *Numer. Heat Transfer A*, vol. 22, pp. 469-485, 1992.

# Legend of Figures

Fig.1

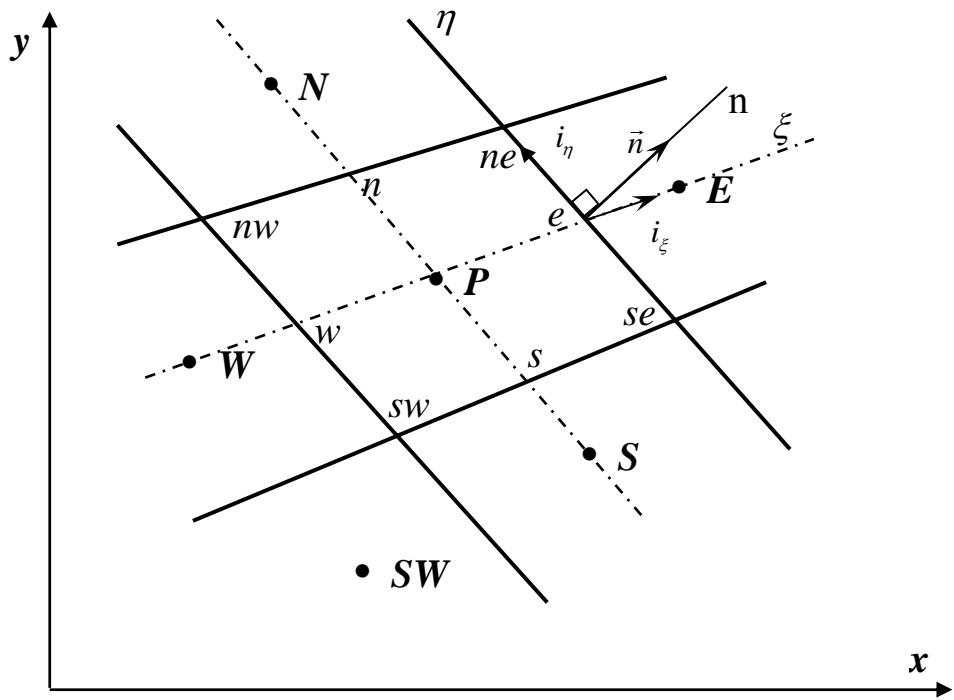

Fig.2

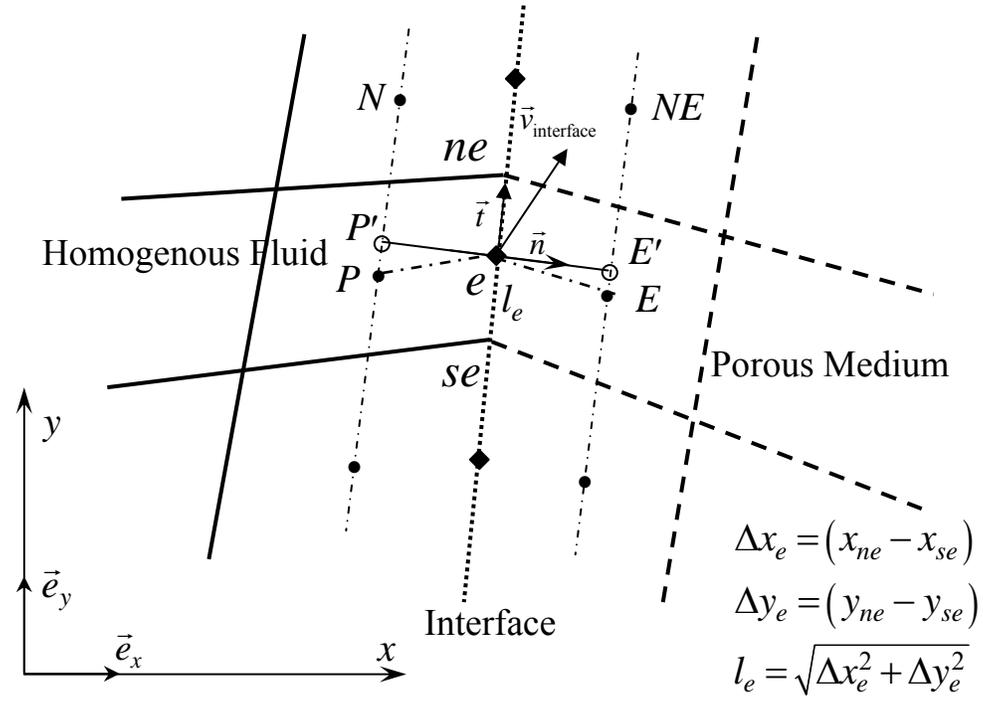

Fig.3

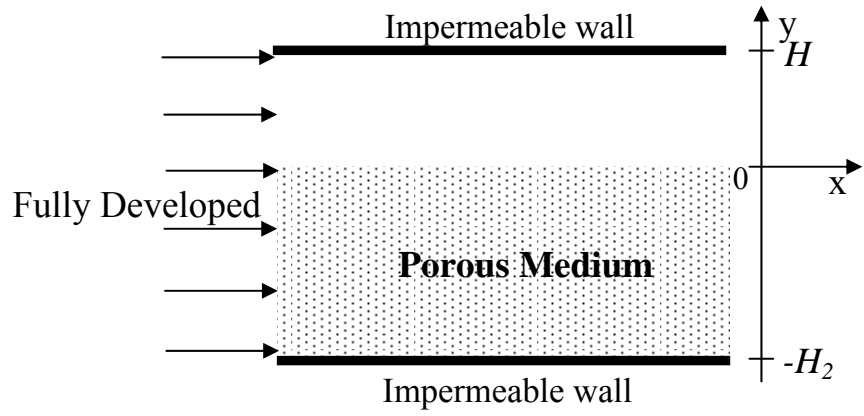

Fig.4

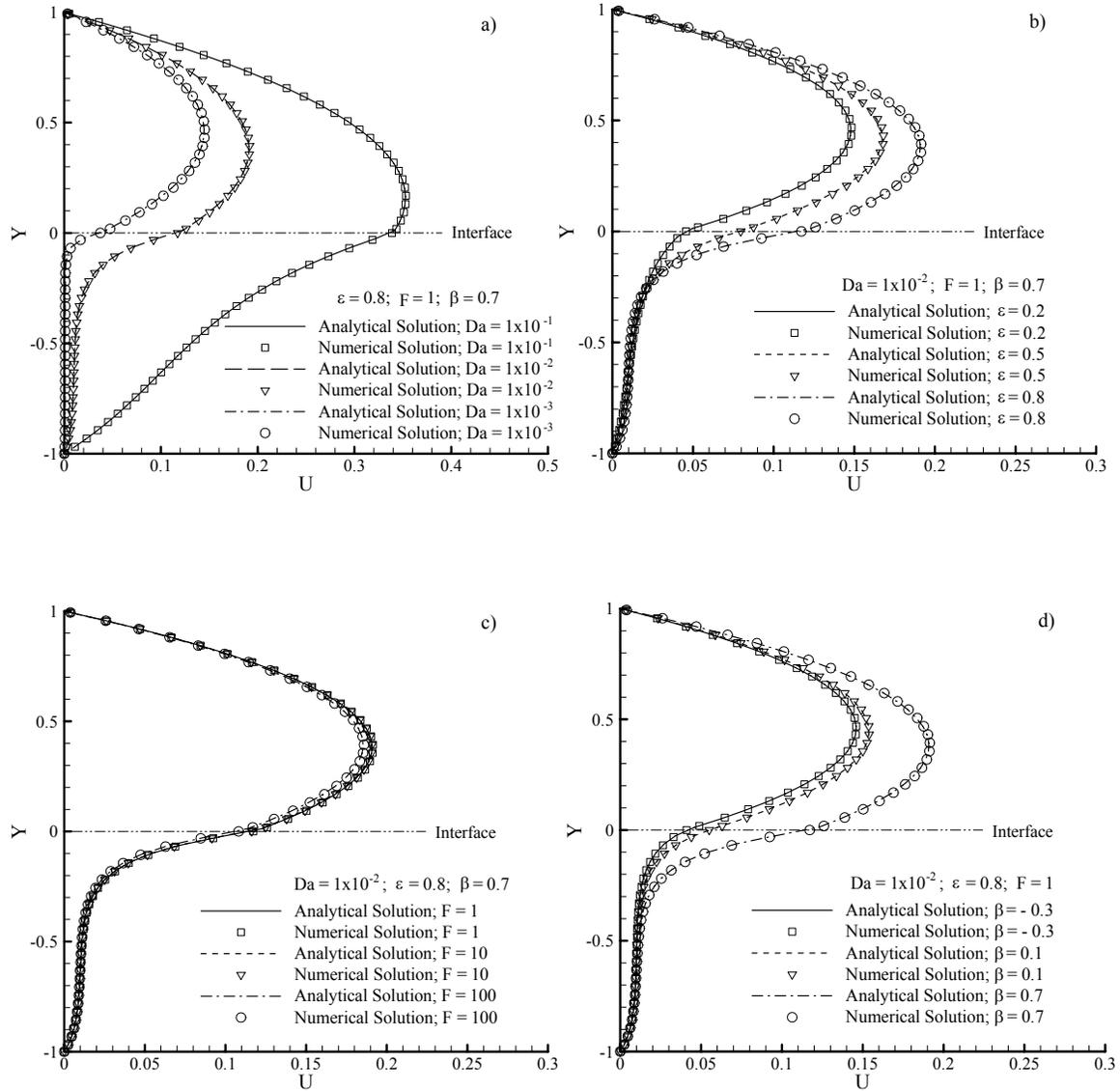

Fig.5

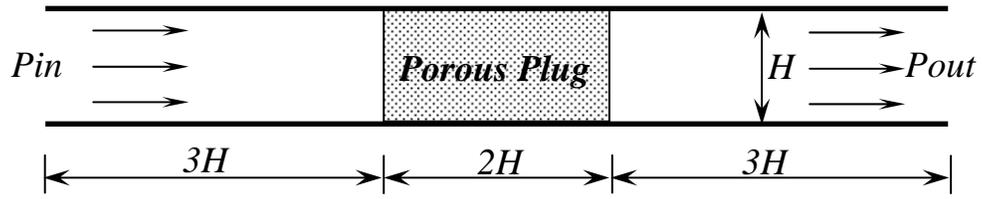

Fig.6

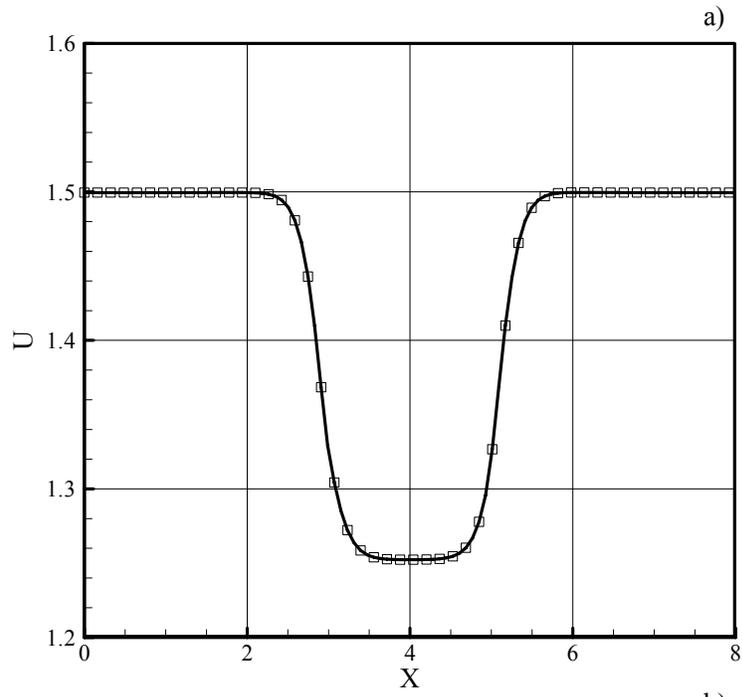

a)

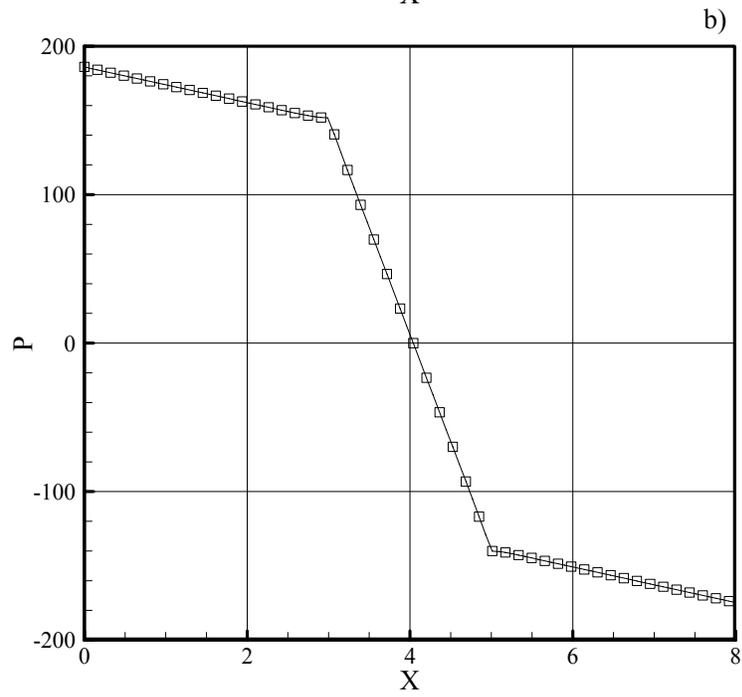

b)

Fig.7

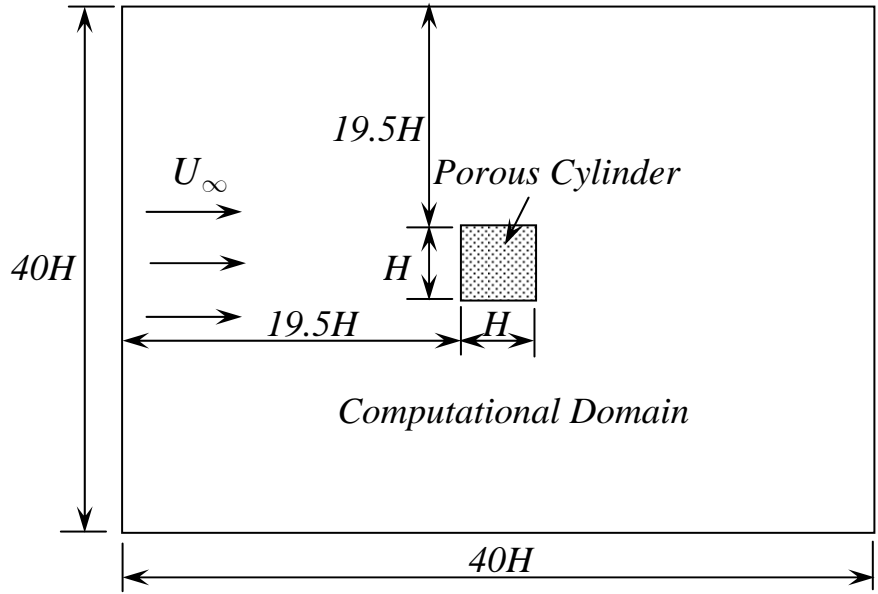

Fig.8

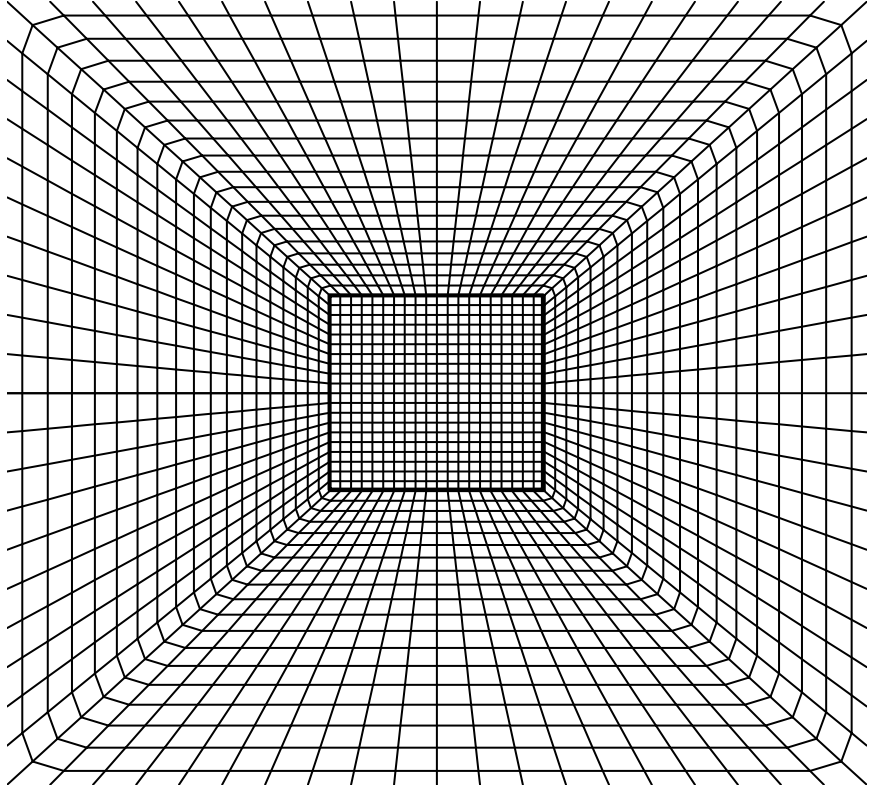

Fig. 9

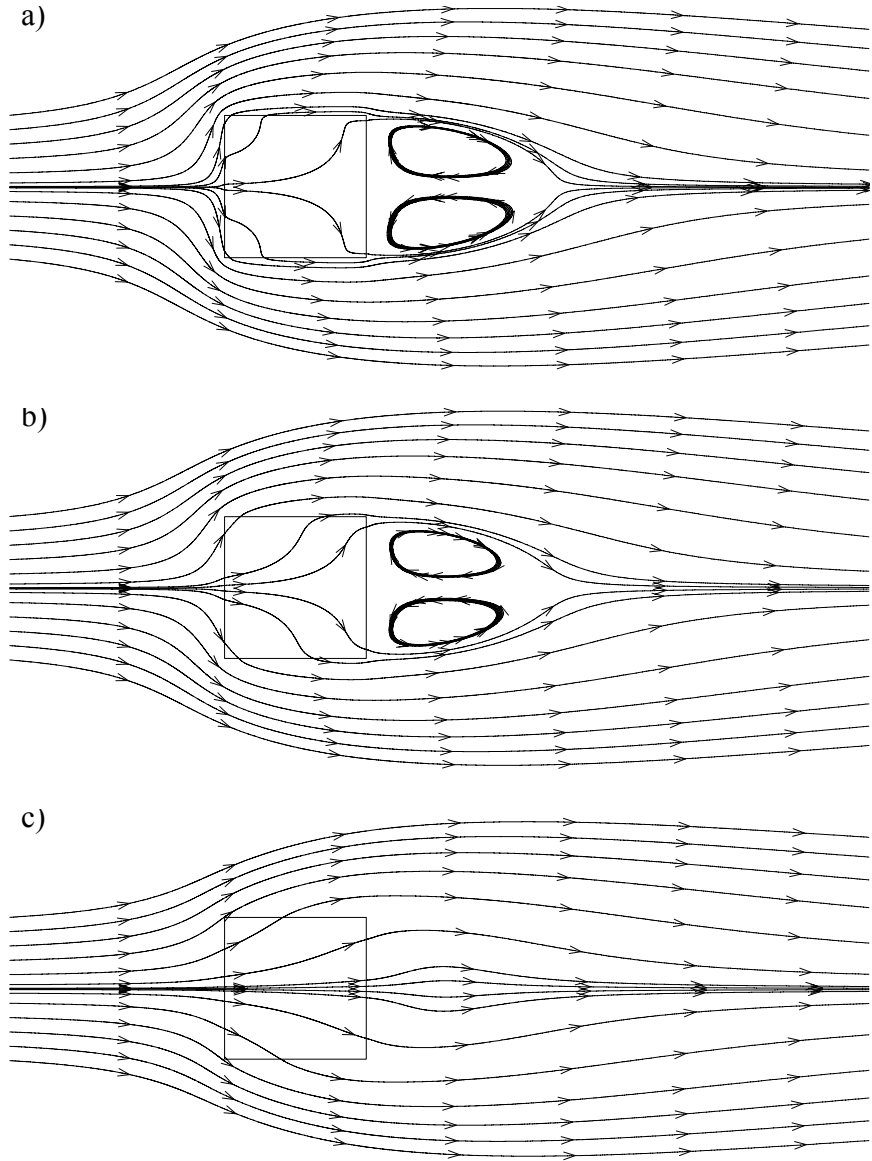